\documentstyle[aps,pra,twocolumn,psfig,floats]{revtex}
\draft
\begin{document}

\title{ \vskip -0.5cm
         \hfill\hfil{\rm\normalsize Printed on \today}\\
         Laser Spinning of Nanotubes: A path to fast-rotating microdevices}

\author{Petr Kr\'al$^{1,2}$ and H. R. Sadeghpour$^{2}$}

\address{$^1$ Department of Chemical Physics, Weizmann Institute of Science,
         76100 Rehovot, Israel}

\address{$^2$ ITAMP, Harvard-Smithsonian Center for Astrophysics,
         Cambridge, Massachusetts 02138}

\maketitle


\begin{abstract}
We show that circularly polarized light can spin nanotubes with GHz 
frequencies. In this method, angular moments of infrared photons are
resonantly transferred to nanotube phonons and passed to 
the tube body by ``umklapp" scattering. We investigate experimental 
realization of this ultrafast rotation in carbon nanotubes, levitating in 
an optical trap and undergoing mechanical vibrations, and discuss 
possible applications to rotating microdevices.
\end{abstract}

\pacs{
32.80.Lg,
62.25.+g,
78.67.Ch,
85.35.Kt
}


Carbon nanotubes \cite{Iijima91} have unique mechanical and electronic 
properties with many potential applications \cite{Dresselhaus}. They
possess a huge Young modulus $Y> 1$ TPa, which adjusts their autonomous 
mechanical oscillations to MHz frequencies \cite{Treacy}.  Moreover, 
their "molecular structures'' remain naturally stable even at large 
deformations \cite{Lourie}.

Therefore, rotationally symmetric structures based on stiff nanotubes 
could form ideal {\it piston-rods} for nanoscale applications.  In 
contrast to chemically driven bio-motors \cite{Caplan}, spinning with 
Hz frequencies, such tubular structures could rotate very fast, if 
angular momentum is efficiently transferred to them and friction is reduced. 

Small heteropolar molecules can be dissociated \cite{Centr}, if 
{\it synchronously} rotated with a dipolar laser trap, which accelerates 
its angular velocity.  Larger molecules \cite{Santamato} and 
micro-particles \cite{Higurashi} can be 
rotated by absorption of angular momentum from circularly polarized or 
"twisted'' laser beams. Nanotubes are excellent candidates for this 
{\it asynchronous} driving, where the system rotational frequency is 
much smaller than the light frequency.  

Here, we investigate ultrafast asynchronous rotation induced in 
nanotubes by excitation of their {\it vibrational} modes with circularly
polarized light. The mode selection is restricted by radiational heating, 
since each photon absorbed by the tube transfers to it angular momentum 
$\hbar$ and energy $\hbar \omega$. The resulting heating can be limited 
in excitation of infrared (IR) $A_{2u}$ or $E_{1u}$ phonon modes, active 
in graphite \cite{Nemanich} and nanotubes \cite{IRmodes}. 

\begin{figure}[t]
\vspace*{-15mm}
\centerline{\psfig{figure=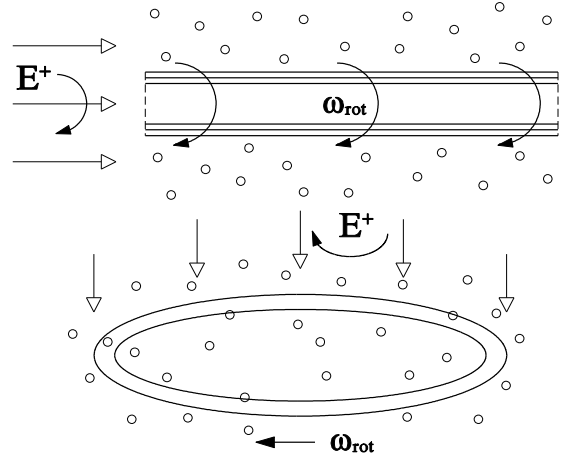,width=1.3\linewidth}}
\caption{Scheme for nanotube (up) and tubular ring (down) spinning with 
angular velocity $\omega_{rot}$ in a laser trap. Their rotation is induced by 
absorption of circularly polarized photons from a laser beam with intensity 
${\bf E}^+$, propagating along the axis of rotational symmetry. Scattering 
of molecules with tubes damps the rotation.}
\label{FIG1}
\end{figure}

\begin{figure}[t]
\vspace*{-25mm}
\centerline{\psfig{figure=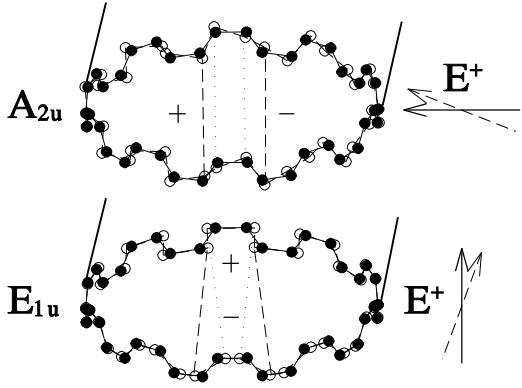,width=2.0\linewidth}}
\vspace*{-45mm}
\caption{Nanotube cross-sections with light-induced atomic displacements 
(open circles) from equilibrium positions (filled circles) in two IR phonon 
modes, with the $A_{2u}$ and $E_{1u}$ symmetries.  A circularly polarized 
light, ${\bf E}^+$, excites phonon waves, which propagate uni-directionally 
on the tube circumference in phase with the light polarization. }
\label{FIG2}
\end{figure}

In Fig.~\ref{FIG1} we show two schemes for spinning nanotubes.  In the upper 
one, circularly polarized light beam propagates along the symmetry axis 
of the single-wall (SWNT) or multi-wall (MWNT) nanotube, levitating in 
an optical trap. The photon angular momentum is transferred to
{\it circularly polarized} phonons, counter-propagating on the tube 
circumference (see Fig.~\ref{FIG2}), and latter passed by scattering to 
the tube body. The angular momentum of light could be also directly passed to 
the nanotube in excitation of its dense rotational levels. The resulting tube 
rotation with angular frequency $\omega_{rot}$ is mostly balanced by friction 
with the surrounding molecules.  In the lower configuration, a closed 
nanotube ring \cite{Liu} is analogously rotated by absorption of 
circularly polarized photons.

We can describe the excitation of nanotube phonons by circularly polarized 
light, and the subsequent relaxation, with the simplified Hamiltonian 
\begin{eqnarray}
H & = & \sum_{\alpha}~ \hbar \omega_{\alpha}~b_{\alpha}^{\dagger}
~b_{\alpha} + \sum_{\alpha_\pm } \mu_{\alpha_\pm} E^{\pm}(t)
\left( b_{\alpha_\pm}^\dagger + b_{\alpha_\pm} \right)
\nonumber \\
& + &
\sum_{\alpha_\pm,\beta,\gamma} \Bigl( c_{\alpha_\pm,\beta,\gamma}
~b_{\alpha_\pm}^{\dagger} b_{\beta} ~b_{\gamma}+H.c. \Bigr)
+ H_{d}\ .
\label{Ham}
\end{eqnarray}
The first two terms describe phonon modes $\alpha=(band,k)$ and 
coupling of the chosen IR circularly polarized optical phonons, with 
operators \cite{Kiselev} $b_{\alpha_\pm}^{\dagger} =2^{-1/2}\, 
(b_{\alpha x}^{\dagger} \pm i b_{\alpha y}^{\dagger})$ and $~b_{\alpha_\pm} 
=2^{-1/2}\, (b_{\alpha x}\mp i b_{\alpha y})$, to the light intensity 
$E^{\pm}(t)$ of the same polarization. The third term denotes decay of 
these IR phonons, with wave vectors $k \approx 0$, into phonon pairs with 
opposite wave vectors $ \pm k$, which most likely come from the same 
acoustical branch \cite{Klemens,Usher}. These also can not carry angular 
quasi-momentum $L$, which is passed to the tube by umklapp processes.
The resulting tube rotation is predominantly damped by scattering with 
molecules, as described in $H_{d}$ \cite{KralDRAG}.

In Fig.~\ref{FIG2}, we show two (doubly degenerate) IR modes in the elementary 
cell, with 40 atoms, of the (10,10) nanotube. In the $A_{2u}$ and $E_{1u}$ 
modes, the atoms move out-of-plane and in-plane, respectively, orthogonal to 
the tube axis \cite{saito}, as shown by open circles. Combination of 
the two degenerate linearly polarized modes forms a {\it circularly 
polarized} phonon mode, of either symmetry, which can absorb angular 
quasi-momentum from circularly polarized photons. The atomic displacements 
break the tube symmetry and induce electric dipoles (+\, -), which follow 
in time the polarization of the circulating electric field ${\bf E}^+$. The
effect does not rely on coherent light and can be also realized in tubular 
rings (see Fig.~\ref{FIG1}).

As an example, we consider excitation of the $A_{2u}$ mode. The total 
number $n_{A_{2u}}^+$ of circularly polarized phonons, excited 
in the vicinity of $k=0$, is given by the Boltzmann equation 
\begin{eqnarray}
\frac{\partial n_{A_{2u}}^+}{\partial t} 
&= & \dot{n}_{A_{2u}}^+ -\frac{n_{A_{2u}}^+-n_{A_{2u};equil.}^+}
{\tau_{A_{2u}}}\ .
\label{Phon}
\end{eqnarray}
$\dot{n}_{A_{2u}}^+$ is their injection rate and $\tau_{A_{2u}}
=2\hbar/ \gamma_{A_{2u}} \approx 2$ ps, is the relaxation time,  
where $\gamma_{A_{2u}} \approx 22$ cm$^{-1}$ is the width of the IR 
phonon lines in nanotubes \cite{IRmodes}. We neglect small populations
$n_{A_{2u}}^-$ of phonons with the opposite polarization, resulting in
scattering. 

The absorption line of the $A_{2u}$ ($E_{1u}$) mode was observed near 
$\omega_{A_{2u}}=870$ cm$^{-1}$ ($1580$ cm$^{-1}$) in both graphite 
and C nanotubes. In graphite, the $A_{2u}$ mode has an oscillator strength 
\cite{Nemanich} $f\approx0.004$, which we assume to approximately hold 
in C nanotubes. Its optical dipole moment is \cite{Boyd} 
$ \mu_{A_{2u}}=e\, \sqrt{3\, \hbar\,  f/ 2\, m_{osc}\,  \omega_{A_{2u}} } 
\approx 10^{-31} \, {\rm Cm}$, 
where $m_{osc}=M_{Carbon}/2$ is the oscillator mass. Using the Fermi's 
Golden rule, and assuming that $n_{A_{2u};equil.}^+\approx 0$, we obtain 
the injection rate
\begin{eqnarray}
\dot{n}_{A_{2u}}^+ \approx \frac{2\pi}{\hbar}~
|\mu_{{A_{2u}}}  E^+|^2\, \rho(\omega_{A_{2u}})\ ,
\label{dotn}
\end{eqnarray}
where $\rho(\omega_{A_{2u}})$ is the density of phonon modes at $k=0$. 
An armchair (10,10) nanotube of length $l=1\, \mu$m has $n\approx 1.6
\times 10^5$ C atoms and $N=n/40=4000$ elementary cells ($A_{2u}$ modes 
with $k\neq 0$). About $10\, \%$ of these modes (around $k=0$)
fall in the energy window $\gamma_{A_{2u}}$, thus giving 
the effective mode density $\rho(\omega_{A_{2u}}) \approx 400/ 
\gamma_{A_{2u}}$.  For a field strength $E^+=10$ kV/m, we then obtain 
from Eq.~\ref{dotn} that $\dot{n}_{A_{2u}}^+ \approx 2.5\times 10^5$ 
s$^{-1}$. The IR phonons thus absorb the angular quasi-momentum with the 
rate $\dot{L}_{A_{2u}} =\hbar \dot{n}_{A_{2u}}^+ \approx 2.5 \times 
10^{-29}$ Nm.

We can understand the angular quasi-momentum umklapp processes by unrolling 
the nanotube, and loosely binding many such sheets into a superlattice of
lattice constant $a_s=2\pi r$, where $r$ is the tube diameter. Then, the
IR phonons modes have the transversal wave vector $K_0=2\pi/a_s$, which
falls in the middle of the second Brillouin mini-zone of size $Q=K_0$.
In a two-phonon umklapp decay, the momentum conservation is $K_0+K_1+K_2=Q$
(transversal wave vectors of the decayed acoustical phonons are $K_{1,2}=0$), 
where the vector $Q$ interconnects centers of the first and second mini-zone.
In the nanotube, we can vector multiply this identity by $\hbar r$, and 
obtain the (umklapp) angular quasi-momentum conservation $L_0+L_1+L_2= \hbar
Q\times r$ ($L_{1,2}=0$), where $L_0=\hbar K_0\times r\equiv \hbar$.

In Eq.~\ref{Phon} these processes are represented by the relaxation 
time $\tau_{A_{2u}}$, which could be derived from Eq.~\ref{Ham} following 
Klemens \cite{Klemens}. Its experimental value, $\tau_{A_{2u}} \approx 2$ ps, 
is in agreement with decay times of the suggested processes realized in 
other systems \cite{Usher}. Since {\it ab initio} calculations of the phonon 
matrix elements $c_{\alpha_\pm,\beta,\gamma}$ are lacking, we use this 
value of $\tau_{A_{2u}}$ in our modeling.  Eq.~\ref{Phon} then gives the 
steady-state angular quasi-momentum in the $A_{2u}$ phonon bath $L_{A_{2u}}= 
\hbar \Delta n_{A_{2u}}^+= \dot{L}_{A_{2u}} {\tau_{A_{2u}}} \approx 5.2 
\times 10^{-41}$ Js.

The angular momentum is transferred to the tube body at the rate $\dot{L} 
\approx \dot{L}_{A_{2u}}$.  Nanotubes in liquids \cite{KralDRAG} or under 
atmospheric conditions would rotate slowly, since collisions with the 
surrounding molecules quickly dissipate the acquired angular momentum.  
On the other hand, in low-vacuum environment, with realistic collisional 
rates $\kappa \approx 10^{-12} - 10^{-13}$ cm$^{3}$ s$^{-1}$, damping 
times of the order $\tau_{damp} \approx 10$ s are readily achievable. 
The tube thus keep a steady-state angular momentum $L \approx 
\dot{L}_{A_{2u}} \tau_{damp} \approx 2.5 \times 10^{-28}$ Js.

The nanotube rotation frequency $\omega_{rot}$ can be found upon 
calculating its principal moments of inertia \cite{Landau,Arnold,Whitley}
\begin{eqnarray}
A=B=M\left(\frac{r_{e}^2+r_{i}^2}{4}+\frac{l^2}{12}\right)\ ,\ \ 
C=M\, \frac{r_{e}^2+r_{i}^2}{2}\ .
\label{ABC}
\end{eqnarray}
Here $M=\rho\, l$, $r_{e}$, $r_{i}$, and $l$ are the nanotube mass ($\rho$ 
is the linear density), exterior and interior radii, and length, respectively. 
For the $(10,10)$ armchair nanotube with $r=(r_{e}+r_{i})/2\approx 0.68$ 
nm and $l=1\, \mu$m, we obtain $M\approx 1.9 \times 10^{-20}$ kg and 
$A\approx 1.6\times 10^{-33}$ kg m$^2 \approx 1.8 \times 10^5\, C$. 

Finally, we find the rotation speed $\omega_{rot}=L/C \approx 28$ GHz
for this elementary nano-mechanical device. Centrifugal acceleration on 
its surface is enormous, $a=r\omega_{rot}^2= 0.5 \times 10^{12}\,  
m/s^2 \approx 10^{11}\, g$. This value surpasses by two orders of 
magnitude the acceleration obtained with sub-millimeter steel balls 
\cite{Beams}, and by five orders of magnitude acceleration in the fastest 
centrifuges \cite{Mashimo}.  Since for 
$a=10^{11}\, g$ the force on each C atom, $F\approx 13\, \mu$eV/\AA\,  
is still negligible with respect to chemical forces (1 eV/\AA), the tube 
rotation could be further increased. On this path to ``tera-gravity",
unique parameters of nanotubes can play a pivotal role.  


We can now discuss in more details practical spinning experiments. Isolated 
SWNT or MWNT have been grown, for example, on an AFM tip \cite{Lieber}, 
which can be later placed inside an optical trap. The nanotube can be 
severed from the tip using, for instance, a focused electron beam 
\cite{ebeam}. Detached tubes could be also transported to the trap by recently 
developed nano-tweezers \cite{tweezers}. The optical trap can be formed 
by two linearly and mutually parallel polarized counter-propagating laser 
beams \cite{Trap}. 

Nonresonant scattering of trap-beam photons from the nanotube with 
polarizability $\alpha$ produces a force, oriented in the direction of 
increasing light intensity $I$, that results in the potential 
$U= -\alpha\, I/2\, c$ ($c$ is the speed of light). The longitudinal
$\alpha_{zz}\approx 500\, \AA^3$/atom and radial $\alpha_{xx}
\approx 25\, \AA^3$/atom static polarizabilities of semiconducting
nanotubes \cite{Benedict} are quite different, and this difference          
is even larger in metallic tubes. Therefore, the tube in the trap remains 
oriented along the beam polarization axis, where it experiences the 
trapping potential
\begin{eqnarray}
U\approx -U_0\, e^{-r^2/\sigma_r^2 -z^2/\sigma_z^2}
\approx -(U_0-S\,r^2)\, e^{-z^2/\sigma_z^2} \ .
\label{well}
\end{eqnarray}
Here $x,y,z$ ($r=x^2+y^2$) are the tube center-of-mass coordinates, and 
$S=S_0\, l=U_0/\sigma_r^2$ is the trap rigidity. To prevent thermal 
escape of the tube from the trap, we consider a trap depth 
$U_0 = n\, \alpha_{zz}\, 
I/2\, c\approx 10$ eV, and obtain $I\approx 1.2$ GW/cm$^2$.  The trap 
laser frequency must be below the band gap, $E_g\approx 1$ eV, and away 
from the frequencies of the tube internal modes, see below. Assuming that 
$\sigma_r\approx 1\, \mu$m, we find $S_0=U_0/l\, \sigma_r^2 \approx 1.6$ 
J/m$^3$.  

Small amounts of defects and adsorbants on the tube walls do not 
prevent its spinning, but can shift its rotation frequency. In 
accordance with the De Laval principle of self-balancing \cite{Whitley}, 
such a partially-coated nanotube floating in the trap would rotate around 
an eccentric axis. Rapid rotation of the nanotube can be also limited by its
mechanical vibrations in the trap, as discussed below. To avoid its large
oscillations, the critical frequencies should be quickly passed during 
the acceleration \cite{Arnold,Whitley}. 

The {\it cylindrical} whirl mode \cite{Whitley} reflects the rigid-body 
vibrations of the tube orthogonal to the trap axis.  The forward 
(backward) cylindrical frequencies are
$\omega_{cyl}=\pm\sqrt{S_0/\rho} \approx \pm 9.2\ {\rm MHz}$.
In the {\it conical} whirl mode, the tube ends move in opposite directions 
with respect to the tube/trap axis. For a tube distorted through the angle 
$\theta$, the torsional moment is $M_F\approx - S_0\, l^3\, \theta/6$, 
resulting in the  Euler's equation, \cite{Arnold} $A\, \omega_{con}^2=C\, 
\omega_{rot}\, \omega_{con} +S_0\, l^3/6$. Using $A\approx \rho\, 
l^3/12$, valid for $l \gg r$, we obtain
\begin{eqnarray}
\omega_{con}=\frac{C \omega_{rot}}{2 A}
\pm \sqrt{ \left(\frac{C \omega_{rot}}{2A}\right)^2
+\frac{2\, S_0}{\rho} }\ .
\label{cowhirl}
\end{eqnarray}
We can see that the modal frequencies depend on $\omega_{rot}$ due to 
{\it gyroscopic} effects \cite{Whitley}. Since the ratio $C/A\approx 
l^{-2}$ is small, the effects are suppressed by the potential $U$, so that 
$\omega_{con} \approx \sqrt{2}\, \omega_{cyl}$. From Eq.~\ref{cowhirl},
we find that they begin to play a role for tube lengths
$l<(r_e+r_i)\, \sqrt{3\, \omega_{rot}/\omega_{cyl}}\approx 130$ nm. 
If the trap is suddenly switched off, a micron-long nanotube rotating with 
frequency $\omega_{rot}=28$ GHz, and initially disturbed on its side, would 
precess with the frequency $\omega_{prec}=C \omega_{rot}/A \approx 175$ kHz.

In long nanotubes, one needs to consider also {\it flexural} vibrations 
\cite{Landau,Arnold,Whitley}. The critical flexural frequencies $\omega_{f}$ 
can be evaluated from the equations for lateral deflections $x(z)$, 
$y(z)$ at different points $z$ along the trap axis, if the rigid body 
approximation is abandoned. The equation for the $x$ deflection is
\begin{eqnarray}
YI~\frac{\partial^4 x}{\partial z^4} & = & 
-\rho\, \frac{\partial^2 x}{\partial t^2}
-S_0~x 
\nonumber \\
& +& a\, \frac{\partial^2 }{\partial t^2} 
\left(\frac{\partial^2 x}{\partial z^2}\right)
+c\,\omega_{rot}\, \frac{\partial }{\partial t} 
\left(\frac{\partial^2 y}{\partial z^2}\right)\ .
\label{FLEX1}
\end{eqnarray}
Here $Y$ is the Young modulus,
$I=\pi\,(r_{e}^4 -r_{i}^4)/4$ is the second moment of nanotube cross-section
and the factors $c=2a=\rho\,(r_{e}^2+r_{i}^2)/2$ are the densities of 
the moments of inertia \cite{Arnold}, which correspond to the bulk 
expressions in Eq.~\ref{ABC} in the limit $l\rightarrow 0$. The equation 
for the $y$ deflection {results} from Eq.~\ref{FLEX1} by exchanging
$x\leftrightarrow y$ and a negative sign in the last term.

The flexural frequencies correspond to the solutions $x=x_0 \cos(\omega_f t)$, 
$y=y_0 \sin(\omega_f t)$ in Eq.~\ref{FLEX1}. This substitution gives an 
ordinary differential equation, identical for both the $x$ and $y$ deflections. 
For simplicity,  we apply the clamped-end approximation, with the boundary 
conditions $x_0(z=\pm \frac{l}{2}) =d^2 x_0(z=\pm \frac{l}{2})/dz^2=0$. The 
solutions are $x_0(z) = A_0\, \cos(\xi z)$ or $x_0(z) = A_0\, \sin(\xi z)$,
where $\xi^2=\bigl(\alpha + \sqrt{\alpha^2-4\,\beta\, YI}\bigr)/2\,YI$,
$\alpha=a\, \omega_f^2-c\, \omega_{rot}\, \omega_f$, and $\beta=S_0-\rho\,
\omega_f^2$.  Therefore, $\xi=n\pi/l$, with $n=1,\, 2,\, 3, ...$ indexing 
the eigenmodes, which leads to the {\it critical} flexural frequencies 
($\omega_{fn} =\omega_{rot}$) 
\begin{eqnarray}
\omega_{fn}=
\sqrt{ \frac{S_0/\rho+ \left( n\pi/l \right)^4\, Y I/\rho}
{1-\left(n\pi r/l \right)^2/2} }\ .
\label{Oflex}
\end{eqnarray}
We use the values $Y\approx 5.5$ TPa and $h=r_{e}-r_{i}=0.066$ nm, found
in molecular dynamics simulations \cite{Yakobson}. 

\begin{figure}[t]
\vspace*{-2mm}
\centerline{\psfig{figure=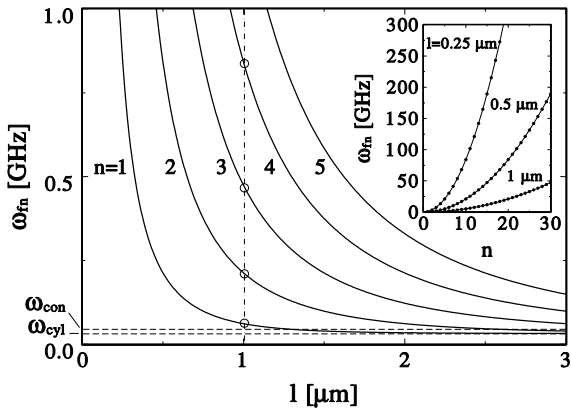,width=1.0\linewidth}}
\vspace*{-2mm}
\caption{Dependence of the critical flexural frequencies $\omega_{fn}$
on the nanotube length $l$. The two thin horizontal dashed lines
correspond to $\omega_{cyl}$ and $\omega_{con}$.
In the inset, we  show $\omega_{fn}$ as a function of
the bending modal number $n$ for nanotubes of different length.}
\label{FIG3}
\end{figure}

In Fig.~\ref{FIG3}, we show the dependence of the lowest critical 
frequencies $\omega_{fn}$ on the tube length $l$, calculated from 
Eq.~\ref{Oflex} using the numerical values for $Y$, $h$, $\rho$, $S_0$ and 
$r$.  For long tubes, the frequencies
$\omega_{fn}$ coincide with $\omega_{cyl}$, while 
for shorter tubes ($l <1.3\, \mu$m), the bending term surpasses the 
trap term, and $\omega_{fn}=\left( n\pi/l \right)^2 \sqrt{Y I/\rho}$.  
In the continuum description, gyroscopic effects become only important 
for high eigenmodes $n\approx l/r$.  In the inset of Fig.~\ref{FIG3}, 
we also show the dependence of $\omega_{fn}$ on $n$ for tubes of different 
lengths. The huge Young modulus $Y$ makes
the density of critical frequencies $\omega_{fn}$ relatively low, 
especially for short nanotubes. This allows for a rapid traversal 
to the ``supercritical state", which is realized above the flexural 
or other vibrational frequencies. 

Rotating nanotubes could form parts of nano-motors, centrifuges or 
stabilizers. Centrifugal studies could be performed inside 
microtubes with large diameters $d\approx 10\ \mu$m \cite{Remskar} or in 
assemblies made from nanotube rings, forming strong but flexible skeletal 
coats. One could also think about possible applications
of rotating tubes in liquids. Slowly rotating coiled nanotubes \cite{Zhang} 
could, for example, propel microscopic systems, which would chemically 
power the rotation of these tubes that attached to their surfaces in bearings 
\cite{Kolmogorov}, as in bio-motors.  We believe that unique properties 
of nanotubes made from carbon and other materials could foster applications 
with rotating micro-elements. 

\vspace{3mm}
\noindent
We would like to thank R. Saito and G. Dresselhaus for data on IR phonon 
modes and several useful discussions. PK would like to acknowledge support from 
EU COCOMO. This work was also supported by the US National Science 
Foundation through a grant to the Institute for Theoretical Atomic and 
Molecular Physics at the Harvard-Smithsonian Center for Astrophysics.



\begin{references}

\bibitem{Iijima91} S.~Iijima, Nature {\bf 354}, 56 (1991).

\bibitem{Dresselhaus} M. S. Dresselhaus, G. Dresselhaus, and P. C. Eklund,
                     {\it Science of Fullerenes and Carbon Nanotubes}
                     (Academic Press Inc., San Diego, 1996).

\bibitem{Treacy} M. M. Treacy, T.W. Ebbesen and J. M. Gibson, Nature {\bf 381},
                 678 (1996); R. Gao {\it et al.}, Phys. Rev. Lett. 
                 {\bf 85}, 622 (2000).

\bibitem{Lourie} O. Lourie, D. M. Cox and H. D. Wagner, Phys. Rev.
                 Lett. {\bf 81}, 1638 (1998).

\bibitem{Caplan} D. Walz and S. R. Caplan, Biophys. J. {\bf 78}, 626 (2000).
               F. A. Samatey {\it et al.}, Nature {\bf 410}, 331 (2001).

\bibitem{Centr} J. Karczmarek, J. Wright, P. Corkum and M. Ivanov, Phys. Rev. 
                Lett. {\bf 82}, 3420 (1999); D. M. Villeneuve {\it et al.},
                Phys. Rev. Lett. {\bf 85}, 542 (2000).

\bibitem{Santamato} E. Santamato, B. Daino, M. Romagnoli, M. Settembre and 
                    Y. R. Shen, Phys. Rev. Lett.  {\bf 57}, 2423 (1986).

\bibitem{Higurashi} E. Higurashi {\it et al.}, Appl. Phys. Lett. {\bf 64}, 
                    2209 (1994); 
                    L. Paterson {\it et al.}, Science {\bf 292}, 912 (2001).

\bibitem{Nemanich} R. J. Nemanich, G. Lucovsky and S. A. Solin, Sol. St.
                 Commun. {\bf 23}, 117 (1977).

\bibitem{IRmodes} J. Kastner {\it et al.}, Chem. Phys. Lett. {\bf 221}, 53 
                (1994);
              U. Kuhlmann, H. Jantoljak, N. Pfander, P. Bernier, C.  Journet 
              and C. Thomsen, Chem. Phys. Lett. {\bf 294}, 237 (1998);

\bibitem{Liu} J.~Liu {\it et al.}, Nature {\bf 385}, 780 (1997).

\bibitem{Kiselev} A. A. Kiselev, Opt. Spectrosc. {\bf 53}, 469 (1982).

\bibitem{Klemens} P. G. Klemens, in {\it Solid State Physics}, ed. by F. Seitz 
                  and D. Turnbull, (Academic Press Inc., NY, 1958), Vol. 7;
                  P. G. Klemens, Phys. Rev. {\bf 148}, 845 (1966).

\bibitem{Usher} S. Usher and G. P. Srivastava, Phys. Rev. B {\bf 50},
                 14179 (1994).

\bibitem{KralDRAG} P. Kr\'al and M. Shapiro, Phys. Rev. Lett. {\bf 86},
                   131 (2001).

\bibitem{saito} R. Saito (private communication).

\bibitem{Boyd} R. W. Boyd, {\it Nonlinear Optics}, (Academic Press, 1992).

\bibitem{Landau} L. D. Landau and E. M. Lifshitz, {\it Elasticity Theory}
                 (Pergamon, Oxford, 1986).

\bibitem{Arnold} R. N. Arnold and L. Maunder, {\it Gyrodynamics and its
                 Enginering Applications}, (Academic, NY/London 1961).

\bibitem{Whitley}  S. Whitley, Rev. Mod. Phys. {\bf 56}, 41 (1984); 
                  {\it ibid} {\bf 56}, 67 (1984). 

\bibitem{Beams}  J. W. Beams, Sci. Am. {\bf 204}, 134 (1961).



\bibitem{Mashimo} T. Mashimo, S. Okazaki and S. Shihabazaki, 
                   Rev. Sci. Instrum. {\bf 67}, 3170 (1996).

\bibitem{Lieber} Ch. L. Cheung, J. H. Hafner, T. W. Odom, K. Kim 
                 and C. M. Lieber, Appl. Phys. Lett. {\bf 76}, 3136 (2000).

\bibitem{ebeam} S. Trasobares {\it et al.}, 
                Eur. Phys. J. B {\bf 22}, 117 (2001).

\bibitem{tweezers} P. Kim and C. M. Lieber, Science {\bf 286}, 2148 (1999).

\bibitem{Trap} A. Ashkin, Phys. Rev. Lett. {\bf 24}, 156 (1970); M. J. Renn,
               R. Pastel and H. J. Lewandowski, Phys. Rev. Lett. {\bf 82}, 
               1574 (1999).

\bibitem{Benedict} L. X. Benedict, S. G. Louie, and M. L. Cohen, Phys. Rev. B 
              {\bf 52}, 8541 (1995).


\bibitem{Yakobson} B. I. Yakobson, C. J. Brabec and J. Bernholc, Phys. 
                Rev. Lett. {\bf 76}, 2511 (1996).



\bibitem{Remskar}  M. Rem\v skar {\it et al.}, Appl. Phys. Lett. {\bf 69},
                 351 (1996).

\bibitem{Zhang} X. B. Zhang {\it et al.}, Europhys. Lett. {\bf 27}, 141 
                 (1994); A. Volodin {\it et al.}, Phys. Rev. Lett. {\bf 84},
                 3342 (2000).

\bibitem{Kolmogorov} A. N. Kolmogorov and V. H. Crespi, Phys. Rev. Lett. 
                   {\bf 85}, 4727 (2000); J. Cumings and A. Zettl, Science
                   {\bf 289}, 602 (2000).


\end{references}
 \end{document}